\documentclass[sigconf]{acmart}

\usepackage{multirow}
\usepackage{float} 
\usepackage{longtable}
\usepackage[table]{xcolor}
\usepackage{amsmath}
\usepackage{amsbsy}
\usepackage{balance}
\usepackage{flushend}
\usepackage{algorithm}
\usepackage{algpseudocode}
\AtBeginDocument{%
  }

\setcopyright{acmlicensed}
\copyrightyear{2025}
\acmYear{2025}
\acmDOI{XXXXXXX.XXXXXXX}
\acmConference[WWW '26]{April 13-17,
  2026}{Dubai, UAE}

\acmISBN{978-1-4503-XXXX-X/2018/06}

\begin{document}

\title{Adaptive Knowledge Transfer for Cross-Disciplinary Cold-Start Knowledge Tracing}

\author{Yulong Deng}
\orcid{1234-5678-9012}


\affiliation{%
  \institution{School of Information Science and Engineering, Yunnan University}
  \city{Kunming}
  \state{Yunnan}
  \country{China}
}
\email{dengyulong@stu.ynu.edu.cn}

\author{Zheng Guan}
\authornotemark[1]

\affiliation{%
  \institution{School of Information Science and Engineering, Yunnan University}
  \city{Kunming}
  \state{Yunnan}
  \country{China}}
\email{larst@affiliation.org}

\author{Min He}
\authornotemark[0]
\authornote{Both authors contributed equally to this research.}

\affiliation{
  \institution{School of Information Science and Engineering, Yunnan University}
  \city{Kunming}
  \state{Yunnan}
  \country{China}}
\email{larst@affiliation.org}

\author{Xue Wang}
\affiliation{
  \institution{Yunnan University}
  \city{Kunming}
  \state{Yunnan}
  \country{China}
}

\author{Jie Liu}
\affiliation{
 \institution{Yunnan University}
 \city{Kunming}
  \state{Yunnan}
  \country{China}}

\author{Zheng Li}
\affiliation{
  \institution{Yunnan University}
  \city{Kunming}
  \state{Yunnan}
  \country{China}}
\renewcommand{\shortauthors}{Deng et al.}

\begin{abstract}
  Cross-Disciplinary Cold-start Knowledge Tracing (CDCKT) faces a critical challenge: insufficient student interaction data in the target discipline prevents effective knowledge state modeling and performance prediction. Existing cross-disciplinary methods rely on overlapping entities between disciplines for knowledge transfer through simple mapping functions, but suffer from two key limitations: (1) overlapping entities are scarce in real-world scenarios, and (2) simple mappings inadequately capture cross-disciplinary knowledge complexity. To overcome these challenges, we propose Mixed of Experts and Adversarial Generative Network-based Cross-disciplinary Cold-start Knowledge Tracing Framework. Our approach consists of three key components: First, we pre-train a source discipline model and cluster student knowledge states into K categories. Second, these cluster attributes guide a mixture-of-experts network through a gating mechanism, serving as a cross-domain mapping bridge. Third, an adversarial discriminator enforces feature separation by pulling same-attribute student features closer while pushing different-attribute features apart, effectively mitigating small-sample limitations. We validate our method's effectiveness across 20 extreme cross-disciplinary cold-start scenarios.
\end{abstract}





\maketitle

\section{Introduction}
Knowledge tracing is a key component of personalized education, which is accomplished by analyzing students' answer records to obtain their knowledge state \cite{abdelrahman2023knowledge, shen2024survey, bai2024survey}. The assessment results can support further customized applications, such as tailored teaching and personalized question recommendations. Therefore, knowledge tracing has attracted great attention in various fields.

As modern education systems evolve rapidly, assessing learners' performance in new subjects has become an important task \cite{akpen2024impact, hung2024evaluation}. Previously, many knowledge tracing models have been developed to improve prediction accuracy. However, many of these models encounter cross-disciplinary cold-start bottlenecks. When a student encounters questions in a new discipline, the model cannot effectively assess the student's performance in the new discipline due to the limitations of data from a single discipline. Therefore, the cross-disciplinary performance of traditional KT has been seriously compromised because they only work in disciplines where students have extensive records. We call this cross-disciplinary cold-start knowledge tracing (CDCKT). CDCKT aims to evaluate a student's potential performance in the target discipline based on their practice records in the source discipline.


There are several researches indicating that relevant methods in recommendation systems have some effect on cross-disciplinary evaluation tasks \cite{gao2023leveraging, hu2023ptadisc}. These strategies primarily focus on how to transfer representations from mature source domains to cold-start target domains \cite{man2017cross, jin2024cross}. They rely on overlapping entities across different domains to capture transfer relationships, thereby obtaining representations usable in the target domain \cite{zang2022survey, zhang2025comprehensive}. However, directly applying these strategies to KT is infeasible, as they overlook two critical questions: First, they employ a single cross-mapping network to handle transfer relationships between different domains \cite{zhu2022personalized}, failing to reflect the characteristic that similar students should share more knowledge transfer patterns \cite{wang2024making}. Second, the number of overlapping entities across different disciplines in real-world scenarios is extremely sparse \cite{guo2023dan}. Consequently, models relying solely on overlapping entities are constrained by sparse supervision signals \cite{liu2024extracting}, unable to fully capture cross-disciplinary relationships.

To address these challenges, we propose Mixed of Experts and Adversarial Generative Network-based Cross-disciplinary Cold-start Knowledge Tracing Framework, a few-shot learning approach for cross-disciplinary knowledge tracing. Specifically, our method approaches the problem from two perspectives. First, \textbf{Overlapping data}. Due to variations in the knowledge state distribution of students across different disciplines, a single mapping function cannot adequately fit cross-domain relationships. To this end, we designed a Mixture of Experts network as a comprehensive cross-domain mapping network. It processes different representations of different students, and different experts perform their respective duties under the guidance of a gated network to capture personalized transfer patterns. Second, \textbf{Complete source domain data}. We leverage complete source domain data to pre-train the model, addressing the sparse supervision problem in the target domain. The pre-trained model extracts student knowledge state representations, which are then clustered into K groups using the silhouette coefficient method \cite{zhang2018differential} to determine the optimal cluster number and K-means \cite{ahmed2020k} for clustering. While  for non-overlapping source domain students (those without counterparts in the target domain), we employ unsupervised feature distribution alignment. Specifically, we enforce that students within the same cluster maintain closer representation distances after cross-domain mapping, while students from different clusters are pushed further apart. This cluster-based constraint enables effective knowledge transfer even without direct entity correspondence between disciplines. Based on the above process, we implemented a generative adversarial network \cite{goodfellow2020generative, aggarwal2021generative} based on generator and discriminator learning. As the generator, the Mixture of Experts mapping network obtains an approximate representation of the student's target discipline. The discriminator distinguishes whether the mapped student representation belongs to the same type. If two students come from the same category, their mapped representations construct a positive sample, and vice versa. The main contributions of this work are summarized as follows:
 
 The contributions of this paper are as follows:
 \begin{itemize}
 \item[$\bullet$]We propose ACKT, a novel framework for cross-disciplinary cold-start knowledge tracing that combines category-guided mapping and adversarial feature alignment, effectively addressing extreme cold-start scenarios where overlapping entities are severely limited.
 \item[$\bullet$]
We design a CMOE mapping network that leverages cluster-level category information to flexibly integrate common cross-disciplinary patterns with personalized transfer preferences, enabling robust knowledge transfer.
 \item[$\bullet$]We introduce an adversarial optimization strategy that exploits non-overlapping entities by enforcing intra-cluster cohesion and inter-cluster separation, effectively mitigating the sparse supervision challenge caused by limited overlapping data.
 \item[$\bullet$]Extensive experiments across 20 cross-disciplinary cold-start scenarios on five real-world datasets demonstrate ACKT's effectiveness and generalizability, with particular improvements in extreme cold-start conditions.
 \end{itemize}


\section{Related work}
\subsection{Knowledge tracing}
Knowledge tracing aims to evaluate students' knowledge states and predict their following performance by leveraging their response records. DKT \cite{piech2015deep} is a pioneering method that introduces deep learning to accomplish KT tasks, utilizing RNNs \cite{fang2021survey} to process input sequences of learning interactions over time. Subsequently developed approaches include Memory-aware KT, Attentive KT, and Graph-based KT \cite{liu2021survey, shen2024survey}. The most representative Memory-aware KT is the Dynamic Key-Value Memory Network (DKVMN) proposed by Zhang et al \cite{zhang2017dynamic}. It incorporates an external memory matrix to store knowledge concepts and update corresponding student mastery levels. Subsequently, Wang proposed a Sequential Key-Value Memory Network (SKVMN) \cite{abdelrahman2019knowledge} to combine the recursive modeling capabilities of DKT with the memory capabilities of DKVMN. As the earlier attention-based KT approach, SAKT \cite{pandey2019self} employs Transformers to capture long-term dependencies among student learning interactions. QAKT \cite{jia2023attentive} automatically learns the Q-matrix \cite{wang2022tracking} from student interactions with the aid of attention mechanisms, thereby freeing itself from the constraints of predefined concept labels. To handle complex graph-related data, graph neural networks have emerged. GKT \cite{nakagawa2019graph} aggregates the temporal knowledge state and embeddings of the concept in the answer and its neighboring concepts. \cite{long2022automatical} is an automatic GKT that utilizes automatic graph self-assessment to measure students' knowledge states without requiring manual annotation. Despite these significant advances, all methods assume the model operates in a single- discipline environment—i.e., using a student's prior performance in discipline A to predict subsequent performance in discipline A. When it is necessary to predict a student's performance in discipline B using interaction records from discipline A, a bottleneck is encountered.

\subsection{Mixture of Experts}
The mixed of expert (MoE) is an advanced machine learning paradigm that enhances model scalability, efficiency, and specialization by dynamically selecting different processing approaches for each input \cite{dimitri2025survey}. In the KT task, distinct experts can focus on different problem patterns, domains, or text representations, leading to more efficient feature extraction. Q-MCKT \cite{zhang2025question} explicitly models students' knowledge acquisition states at both question and concept levels, employing the mixed of expert technique to capture more robust and precise acquisition states for prediction support. RouterKT \cite{liao2025routerkt} treats multi-head attention as experts and designs a routing mechanism that encourages experts to focus on context-aware interaction dependencies and temporal dynamics. In this paper, we regard experts as bridges carrying cross-disciplinary information, enabling the model to capture transmission patterns of shared preferences.

\begin{figure*}
	\centering
	\includegraphics[width=\linewidth]{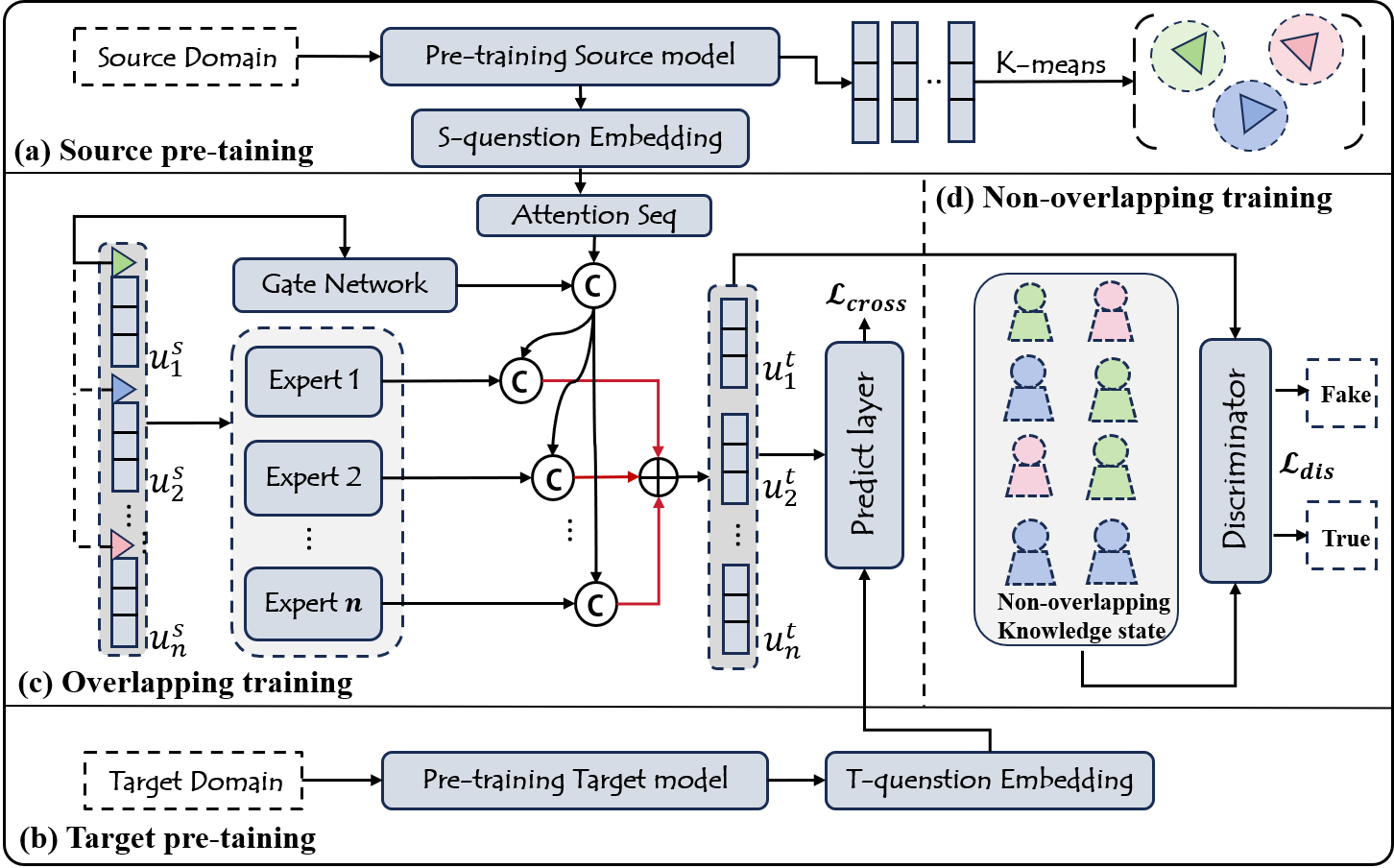}
	\caption{The overview Architecture of ACKT. It consists of two stages, where stage 1 is to extract high-quality knowledge states and question embeddings, and to cluster the knowledge states, as shown in (a), (b). Stage 2 mainly consists of a category-enhanced mixture of experts mapping module (CMOE) and unsupervised preference distribution adversarial optimization, as shown in (c), (d).}
	\label{FIG:2}
\end{figure*}

\section{Method}
In this section, we propose a cross-disciplinary cold-start framework called ACKT. First, we present the definition of cross-disciplinary cold-start knowledge tracing. Second, we describe the pre-training backbone and knowledge state clustering. Third, we introduce the proposed Mixture of Experts mapping network. Fourth, we introduce a generative adversarial method to incorporate non-overlapping student data into the model training. Finally, we review the overall training process of the model. Figure 2 illustrates the overall architecture of our method.
\subsection{Problem formulation}
In cross-disciplinary cold-start knowledge tracing, we have a source discipline and a target discipline. Each discipline has a set of students $U=\{{u}_1,u_2,\ldots\}$, a set of questions $Q=\{q_1,\ q_2,\ \ldots\}$, a set of concepts $C=\{c_1,\ c_2,\ \ldots\}$, and a set of scores $R=\{{r}_1,r_2,\ldots\}$. The source discipline has $n_s$ users, $m_s$ questions, and $k_s$ concepts, while the target discipline has $n_t$ users, $m_t$ questions, and $k_t$ concepts. To distinguish between the two disciplines, we use $U_s, Q_s, C_s$, and $R_s$ to represent the student set, question set, concept set, and score set of the source discipline, and $U_t, Q_t, C_t$, and $R_t$ to represent the student set, question set, concept set, and score set of the target discipline. We define the overlapping students between the two disciplines as $U_o=U_s\cap U_t$.
Given the above data, our goal is to predict the possible performance of unknown non-overlapping disciplines in the target discipline: ${\hat{p}}_{cross}=P\left(r_{n+1}^t=1\middle| u_i^s,q_{n+1}^t\right)$, where $u_i^s\in U_s\notin U_o$ and $q_j^t$ represents the questions in the target discipline.

\subsection{Pre-training and Clustering of Source Disciplines}
During the pre-training phase, our goal is to obtain different categories of student knowledge states from resource-rich source disciplines and train an embedding layer that can quickly adapt to resource-poor target disciplines.
\subsubsection{KT backbone}
We use existing knowledge tracing models as the backbone structure for the pre-training phase. As shown in figure 2, most KT models require knowledge states to be concatenated with question representations before prediction, while our framework relies on the same pre-training representations, and thus our framework can be well adapted to the majority KT models. For backbone settings, please refer to \cite{xie2024domain}.

\begin{figure}[h]
	\centering
	\includegraphics[width=\linewidth]{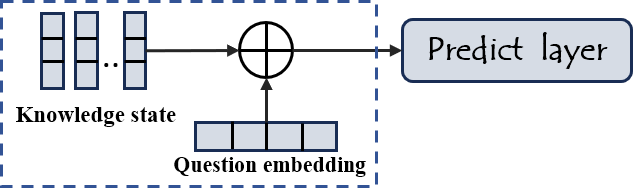}
	\caption{Feature extraction process in the prediction phase for general knowledge tracing methods}
	\label{FIG:1}
\end{figure}

\subsubsection{Clustering of source knowledge states}
For more efficient classification of students' knowledge states into different categories, we use the MiniBatchKMeans \cite{hicks2021mbkmeans} method to cluster the knowledge states of all students. Since manually determining the optimal number of clusters K is time-consuming and labor-intensive, we set $K\in[2,K_{max}]$ and then calculate the most appropriate K value using the contour coefficient method. Specifically, as follows:
\begin{equation}
    \gamma(i)=\frac{\alpha(i)-\beta(i)}{max\{\alpha(i),\beta(i)\}}\label{1}
\end{equation}
where $\alpha(i)$ represents the average distance between sample i and other samples in the same cluster (intra-cluster distance). $\beta(i)$ represents the average distance between sample i and all samples in the nearest cluster (excluding its own cluster) (inter-cluster distance). The final contour coefficient is determined by the average value of all sample contour coefficients:
\begin{equation}
    Silhouette\ Score=\frac{1}{n}\sum_{i=1}^{n}{\gamma(i)}\label{2}
\end{equation}

Without loss of generality, we use $U_o$, $U_s$, and $U_t$ to represent the knowledge states of overlapping students, source students, and target students, respectively. We assign these users to $K$ categories by maximizing the conditional probability:
\begin{equation}
    \mathbf{c}_i = \arg\max_k p\left(\mathbf{c}^k \mid u_i^s\right)\label{3}
\end{equation}
where $\mathbf{c}^k$ denotes the $k$th category, $u_i^s$ denotes the $i$th source student, and $p\left(\mathbf{c}^k\middle| u_i^s\right)$ denotes the probability that the ith user belongs to category k. This category information will play an important role in the subsequent discussion.
\subsection{Category-enabled Mixture-of-Experts Mapping}
Upon obtaining category information for each student, we model each student's unique characteristics, including interaction sequence characteristics and personalized knowledge preferences. Such as Different learning records contribute differently to cross-domain transfer \cite{otovic2022intra}. We use simple but effective attention mechanisms to obtain the interaction sequence features of each student. Without loss of generality, let $C_i=\{c_1^s,\ c_2^s,\ c_3^s,...,c_n^s\}$ represent the interaction sequence of student $i$ in the source domain, where n represents the length of the interaction sequence. The attention network is formulated as follows:
\begin{equation}
{\bar{a}}_j^s=att\left(c_j^s;\theta_a\right)\label{4}
\end{equation}
\begin{equation}
a_j^s=\frac{exp{\left({\bar{a}}_j^s\right)}}{\sum_{c_k^s\in\mathcal{S}_{i}} e x p{\left({\bar{a}}_k^s\right)}}\label{5}
\end{equation}
where $att(\cdot)$ denotes an attention network comprising two layers of fed-forward networks, and $\theta_a$ denotes the parameters of the attention network. ${\bar{a}}_j^s$ is the normalized attention score for $c_j^s$, which can be interpreted as the importance of $c_j^s$ in the personalized mapping. Then, the user's interaction sequence is weighted and summed to obtain the user feature representation.
\begin{equation}
\mathbf{p}_\mathbf{i}=\sum_{c_k^s\in\mathcal{S}_{i}}\label{6}{a_j^s\mathbf{c}_\mathbf{k}^\mathbf{s}}
\end{equation}
where $\mathbf{p}_\mathbf{i}\in\mathbb{R}^d$ represents the transferable feature embedding of student i.

Next, we extract the pre-trained student representation and student category representation, and fuse them with the feature representation of the interaction sequence as expert input to learn each expert's output. The process is formalized as follows:
\begin{equation}
\mathbf{z}_\mathbf{i}=\mathbf{u}_\mathbf{i}^\mathbf{s}\oplus\mathbf{p}_\mathbf{i}\oplus\mathbf{c}_\mathbf{i}\label{7}
\end{equation}
where $\mathbf{z}_\mathbf{i}\in\mathbb{R}^\mathbf{d}$ represents student i's input to the expert network. The specific form of the Mixture of Experts mapping function is defined as follows:
\begin{equation}
G\left(Z\right)=\mathrm{Softmax}\left(\textbf{W}_gZ+\textbf{b}_g\right)\label{8}
\end{equation}
\begin{equation}
F_{\mathrm{expert}}\left(C\right)=\sigma\left(\textbf{W}_hC+\textbf{b}_h\right)\label{9}
\end{equation}
\begin{equation}
{\hat{u}}_i^t=\sum_{x=1}^{X}{G_x\left(c_i\right)}\cdot F_x\left(z_i\right)\label{10}
\end{equation}
where $G\left(\mathbf{Z}\right)$ and $F_{\mathrm{expert}}\left(\mathbf{C}\right)$ denote the formulas for the gate and expert networks, respectively, both of which consist of multiple linear layers and activation functions. The category representation C of the source domain student is the input to the gate network. $\mathbf{W}_{g}$ and $\mathbf{b}_{g}$ are the feature transformation matrix and bias matrix, respectively, while $\mathbf{W}_{h}$ and $\mathbf{b}_{h}$ have similar definitions in the expert network. $\sigma$ is the activation function, and X is a hyperparameter representing the number of experts. ${\hat{u}}_i^t$ is the mapping target representation we obtain, which reflects the approximate student target representation.

\subsection{Preference Distribution Adversarial}
After obtaining approximate student preferences in the target discipline, a common method is to optimize the model by aligning the true labels of students in the target discipline. However, extreme cold start environments limit the number of labels, preventing the model from fully learning to capture cross-disciplinary relationships.
To address this, we propose category preference alignment optimization based on equal variation learning, which assumes that students with similar performance in the source discipline will also have similar performance in the target discipline. Specifically, drawing inspiration from adversarial generative networks, we treat ${\hat{u}}_i^t$ as the generator's output and use a discriminator to judge whether overlapping students and non-overlapping students belong to the same category, where the number of non-overlapping students is obtained through sampling, which we set to $N_s$. We will detail the settings and optimization methods for generators and discriminators in the following sections. 

\textbf{Generator:} The generator need sample students of the same type as us a and non-overlapping students of different types. For simplicity, we assume $N_s$ = 1 at this point.
\begin{align}\left.\left\{
\begin{array}
{c}\mathbf{z}=\mathbf{u}^s\oplus\mathbf{p}\oplus\mathbf{c}, \\
\hat{u}_a^t=\mathrm{CMOE}(\mathbf{z}_a,\mathbf{c}_a) \\
\hat{u}_b^t=\mathrm{CMOE}(\mathbf{z}_b,\mathbf{c}_b),\quad c_b=c_a \\
\hat{u}_c^t=\mathrm{CMOE}(\mathbf{z}_c,\mathbf{c}_c),\quad c_c\neq c_a,\label{11}
\end{array}\right.\right.\end{align}
where $u^s, p, c$ denote the pre-trained knowledge state, concept sequence representation, and category representation, respectively; $u^s_b$ shares the same category as $u^s_a$, while $u^s_c$ represents a different category sample. Through the CMOE module, overlapping student $u^s_a$, same-category source student $u^s_b$, and different-category source student $u^s_c$ can be jointly mapped to the target discipline, and corresponding knowledge states $\hat{u}^t_a, \hat{u}^t_b$, and $\hat{u}^t_c$ are obtained.

\textbf{Discriminator:} Given the student knowledge state representations after mapping, we concatenate each pair of students' mapped representations as input to the discriminator D. For two students sharing the same category, the discriminator outputs a “true” value; otherwise, it outputs a “false” value. The optimization is defined as follows:
\begin{align}
\mathcal{L}_{\mathrm{dis}}
&= \min_{\theta_d}\;
\mathbb{E}_{\substack{u_a^s\sim \mathcal{U}^o\\ u_c^s\notin c_a}}
\log\mathcal{D}\!\left(\hat{u}_a^t \Vert \hat{u}_c^t\right)
\nonumber\\[2pt]
&\quad -\;
\mathbb{E}_{\substack{u_a^s\sim \mathcal{U}^o\\ u_b^s\in c_a}}
\log\mathcal{D}\!\left(\hat{u}_a^t \Vert \hat{u}_b^t\right).\label{12}
\end{align}
In practice, we employ a three-layer MLP to implement the discriminator D, where $\theta_d$ represents its parameters. This approach enables students from non-overlapping segments to participate in model training, thereby effectively mitigating the sparse supervision caused by extreme cold-start challenges.

\subsection{Overall training process of the model}

The complete training process is shown in Algorithm 1. In which, The optimization of the model is divided into two parts: first, supervised optimization of the model through overlapping student interaction data labels; second, unsupervised optimization of the adversarial loss of the adversarial generation part. For the first part, we use cross-entropy as the loss function. It is defined as follows:
\begin{equation}
    \text{$\mathcal{L}$}_{cross} = \sum \left( r_l^{t} \log(\widehat{p}^t_l) + (1 - r_l^{t}) \log(1 - \widehat{p}^t_l) \right)\label{13}
\end{equation}
In addition to cross-entropy loss, we incorporate our proposed generative adversarial loss to train the model using non-overlapping students, the final loss function is as follows:
\begin{equation}\label{14}
\mathcal{L}=\mathcal{L}_{cross}+\lambda \mathcal{L}_{dis}    
\end{equation}

\begin{algorithm}[H]
\caption{Overall training process of ACKT}  \label{al}
\begin{algorithmic}
\State \textbf{Input:} $U_s,U_t,U_o,Q_s,C_s,Q_t,C_t$
\State \textbf{Fixed Parameters:} \\$\lambda=200,N_s=16,n=24$\\Pre-train and Train epoch: $K_p \& K  $ 
\State Batch size: 64; Initial learning rate: 0.001
\State \textbf{Perform pre-training}
\end{algorithmic}
\begin{algorithmic}[1]
\For{$k_p$ \textbf{in} range($K_P$)}
    \State  Update pre-train model parameters
\EndFor
\end{algorithmic}
\begin{algorithmic}
\State \textbf{K-means clustering}
\end{algorithmic}
\begin{algorithmic}[1]
\State Extracting Knowledge States from Pre-trained Model
\State Computing student categories using K-means Eq.(\ref{1})-(\ref{3}).
\end{algorithmic}

\begin{algorithmic}
\State \textbf{Perform training}
\end{algorithmic}

\begin{algorithmic}[1]
\For{$k$ \textbf{in} range($K$)}
    \State  Computing interactive sequence features via Eq.(\ref{4})-(\ref{6})
    \State Accessing students' discipline preference via Eq.(\ref{7})-(\ref{11}) 
    \State Calculate unsupervised discriminator loss via Eq.(\ref{12})
    \State Calculate cross-disciplinary prediction loss via Eq.(\ref{13})
    \State Calculate total loss via Eq.(\ref{14})
    \State  Update model parameters
\EndFor
\end{algorithmic}
\end{algorithm}

\begin{table*}[]
\caption{Performance comparison. The best cross-disciplinary performance is highlighted in bold.}
\label{tab:my-table}
\begin{tabular}{
>{\columncolor[HTML]{FFFFFF}}c 
>{\columncolor[HTML]{FFFFFF}}c 
>{\columncolor[HTML]{FFFFFF}}c 
>{\columncolor[HTML]{FFFFFF}}c 
>{\columncolor[HTML]{FFFFFF}}c 
>{\columncolor[HTML]{FFFFFF}}c 
>{\columncolor[HTML]{FFFFFF}}c 
>{\columncolor[HTML]{FFFFFF}}c 
>{\columncolor[HTML]{FFFFFF}}c 
>{\columncolor[HTML]{FFFFFF}}c 
>{\columncolor[HTML]{FFFFFF}}c 
>{\columncolor[HTML]{FFFFFF}}c 
>{\columncolor[HTML]{FFFFFF}}c }
\hline
\cellcolor[HTML]{FFFFFF}                                                 & \multicolumn{3}{c}{\cellcolor[HTML]{FFFFFF}\textbf{Java-Python}} & \multicolumn{3}{c}{\cellcolor[HTML]{FFFFFF}\textbf{Java-C}}     & \multicolumn{3}{c}{\cellcolor[HTML]{FFFFFF}\textbf{Java-C++}}   & \multicolumn{3}{c}{\cellcolor[HTML]{FFFFFF}\textbf{Java-Ds}}   \\ \cline{2-13} 
\multirow{-2}{*}{\cellcolor[HTML]{FFFFFF}Methods}                        & \textbf{AUC}         & \textbf{ACC}        & \textbf{RMSE}       & \textbf{AUC}         & \textbf{ACC}         & \textbf{RMSE}     & \textbf{AUC}         & \textbf{ACC}        & \textbf{RMSE}      & \textbf{AUC}        & \textbf{ACC}        & \textbf{RMSE}      \\ \hline
DisKT                                                                    & 0.5998               & 0.7359              & 0.4377              & 0.5488               & 0.5580               & 0.6023            & 0.5517               & 0.5351              & 0.6260             & 0.5545              & 0.5324              & 0.6307             \\
RouterKT                                                                 & 0.6181               & 0.6972              & 0.4784              & 0.5936               & 0.6778               & 0.4903            & 0.6149               & 0.6776              & 0.4940             & 0.6109              & 0.6952              & 0.4821             \\
csKT                                                                     & 0.5885               & 0.5900              & 0.5541              & 0.5520               & 0.5290               & 0.5962            & 0.5654               & 0.5794              & 0.5716             & 0.5605              & 0.5552              & 0.5961             \\
CL4KT                                                                    & 0.5922               & 0.7278              & 0.4612              & 0.5735               & 0.7130               & 0.4686            & 0.5932               & 0.7177              & 0.4696             & 0.5972              & 0.7314              & 0.4559             \\
sparseKT                                                                 & 0.5852               & 0.5803              & 0.6478              & 0.5465               & 0.5284               & 0.6868            & 0.5574               & 0.5773              & 0.6501             & 0.5574              & 0.5773              & 0.6501             \\
Ours                                                                     & \textbf{0.7065}      & \textbf{0.7597}     & \textbf{0.4083}     & \textbf{0.6870}      & \textbf{0.7391}      & \textbf{0.4263}   & \textbf{0.6865}      & \textbf{0.7367}     & \textbf{0.4379}    & \textbf{0.6694}     & \textbf{0.7667}     & \textbf{0.4097}    \\ \hline
\multicolumn{1}{l}{\cellcolor[HTML]{FFFFFF}}                             & \multicolumn{3}{c}{\cellcolor[HTML]{FFFFFF}\textbf{Python-Java}} & \multicolumn{3}{c}{\cellcolor[HTML]{FFFFFF}\textbf{Python-C}}   & \multicolumn{3}{c}{\cellcolor[HTML]{FFFFFF}\textbf{Python-C++}} & \multicolumn{3}{c}{\cellcolor[HTML]{FFFFFF}\textbf{Python-Ds}} \\ \cline{2-13} 
\multicolumn{1}{l}{\multirow{-2}{*}{\cellcolor[HTML]{FFFFFF}Methods}}    & \textbf{AUC}         & \textbf{ACC}        & \textbf{RMSE}       & \textbf{AUC}         & \textbf{ACC}         & \textbf{RMSE}     & \textbf{AUC}         & \textbf{ACC}        & \textbf{RMSE}      & \textbf{AUC}        & \textbf{ACC}        & \textbf{RMSE}      \\ \hline
DisKT                                                                    & 0.5627               & 0.5391              & 0.6560              & 0.5355               & 0.5515               & 0.6324            & 0.5688               & 0.6024              & 0.6020             & 0.5220              & 0.5259              & 0.6545             \\
RouterKT                                                                 & 0.6129               & 0.6937              & 0.4988              & 0.582                & 0.6798               & 0.5095            & 0.5887               & 0.6803              & 0.515              & 0.5880              & 0.6901              & 0.5012             \\
csKT                                                                     & 0.5676               & 0.5674              & 0.6067              & 0.5371               & 0.5248               & 0.6277            & 0.5391               & 0.5389              & 0.6159             & 0.5450              & 0.5419              & 0.6118             \\
CL4KT                                                                    & 0.6113               & 0.7357              & 0.4494              & 0.5781               & 0.7158               & 0.4657            & 0.5828               & 0.6988              & 0.4734             & 0.5744              & 0.6978              & 0.4792             \\
sparseKT                                                                 & 0.5814               & 0.6098              & 0.6247              & 0.5434               & 0.5495               & 0.6712            & 0.5441               & 0.5640              & 0.6603             & 0.5489              & 0.5657              & 0.6590             \\
Ours                                                                     & \textbf{0.6886}      & \textbf{0.7547}     & \textbf{0.4107}     & \textbf{0.6827}      & \textbf{0.7438}      & \textbf{0.4313}   & \textbf{0.6718}      & \textbf{0.7362}     & \textbf{0.4428}    & \textbf{0.6680}     & \textbf{0.7588}     & \textbf{0.4113}    \\ \hline
\cellcolor[HTML]{FFFFFF}{\color[HTML]{000000} }                          & \multicolumn{3}{c}{\cellcolor[HTML]{FFFFFF}\textbf{C-Java}}      & \multicolumn{3}{c}{\cellcolor[HTML]{FFFFFF}\textbf{C-Python}}   & \multicolumn{3}{c}{\cellcolor[HTML]{FFFFFF}\textbf{C-C++}}      & \multicolumn{3}{c}{\cellcolor[HTML]{FFFFFF}\textbf{C-DS}}      \\ \cline{2-13} 
\multirow{-2}{*}{\cellcolor[HTML]{FFFFFF}{\color[HTML]{000000} Methods}} & \textbf{AUC}         & \textbf{ACC}        & \textbf{RMSE}       & \textbf{AUC}         & \textbf{ACC}         & \textbf{RMSE}     & \textbf{AUC}         & \textbf{ACC}        & \textbf{RMSE}      & \textbf{AUC}        & \textbf{ACC}        & \textbf{RMSE}      \\ \hline
DisKT                                                                    & 0.5545               & 0.5783              & 0.5996              & 0.5360               & 0.5652               & 0.6021            & 0.5344               & 0.5777              & 0.6123             & 0.5532              & 0.5857              & 0.5869             \\
RouterKT                                                                 & 0.6132               & 0.7039              & 0.4739              & 0.5999               & 0.6806               & 0.5029            & 0.6177               & 0.6849              & 0.4836             & 0.6042              & 0.6957              & 0.4916             \\
csKT                                                                     & 0.5981               & 0.6559              & 0.5072              & 0.5986               & 0.6604               & 0.5063            & 0.6104               & 0.6796              & 0.4866             & 0.5833              & 0.6557              & 0.5032             \\
CL4KT                                                                    & 0.6262               & 0.7307              & 0.4485              & 0.6096               & 0.6950               & 0.4709            & 0.6036               & 0.6964              & 0.4731             &       0.6054              & 0.7163              & 0.4590                   \\
sparseKT                                                                 & 0.6254               & 0.7063              & 0.5419              & 0.6141               & 0.6947               & 0.5525            & 0.6178               & 0.7131              & 0.5357             & 0.6081              & 0.7244              & 0.5250             \\
Ours                                                                     & \textbf{0.6916}      & \textbf{0.7667}     & \textbf{0.4054}     & \textbf{0.7057}      & \textbf{0.7678}      & \textbf{0.4091}   & \textbf{0.6854}      & \textbf{0.7405}     & \textbf{0.4303}    & \textbf{0.6736}     & \textbf{0.7842}     & \textbf{0.4009}    \\ \hline
\multicolumn{1}{l}{\cellcolor[HTML]{FFFFFF}}                             & \multicolumn{3}{c}{\cellcolor[HTML]{FFFFFF}\textbf{C++-Java}}    & \multicolumn{3}{c}{\cellcolor[HTML]{FFFFFF}\textbf{C++-Python}} & \multicolumn{3}{c}{\cellcolor[HTML]{FFFFFF}\textbf{C++-C}}      & \multicolumn{3}{c}{\cellcolor[HTML]{FFFFFF}\textbf{C++-DS}}    \\ \cline{2-13} 
\multicolumn{1}{l}{\multirow{-2}{*}{\cellcolor[HTML]{FFFFFF}Methods}}    & \textbf{AUC}         & \textbf{ACC}        & \textbf{RMSE}       & \textbf{AUC}         & \textbf{ACC}         & \textbf{RMSE}     & \textbf{AUC}         & \textbf{ACC}        & \textbf{RMSE}      & \textbf{AUC}        & \textbf{ACC}        & \textbf{RMSE}      \\ \hline
DisKT                                                                    & 0.5487               & 0.6083              & 0.5875              & 0.5639               & 0.5802               & 0.6026            & 0.5350               & 0.6002              & 0.5879             & 0.5430              & 0.5116              & 0.6430             \\
RouterKT                                                                 & 0.6098               & 0.6958              & 0.4845              & 0.5969               & 0.6739               & 0.4961            & 0.6018               & 0.6717              & 0.4955             & 0.5944              & 0.6873              & 0.4881             \\
csKT                                                                     & 0.5968               & 0.5996              & 0.5552              & 0.5707               & 0.5795               & 0.5808            & 0.5657               & 0.5851              & 0.5952             & 0.5676              & 0.5874              & 0.5743             \\
CL4KT                                                                    & 0.6085               & 0.6840              & 0.4725              &   0.5976               & 0.7058               & 0.4666            &       0.5995               & 0.7079              & 0.4628             &       0.6015              & 0.7409              & 0.4478                   \\
sparseKT                                                                 & 0.6095               & 0.6723              & 0.5724              & 0.5758               & 0.6103               & 0.6243            & 0.5714               & 0.6028              & 0.6303             & 0.5702              & 0.6329              & 0.6059             \\
Ours                                                                     & \textbf{0.7006}      & \textbf{0.7852}     & \textbf{0.4002}     & \textbf{0.6978}      & \textbf{0.7478}      & \textbf{0.4161}   & \textbf{0.6910}      & \textbf{0.7361}     & \textbf{0.4268}    & \textbf{0.6621}     & \textbf{0.7666}     & \textbf{0.4105}    \\ \hline
\multicolumn{1}{l}{\cellcolor[HTML]{FFFFFF}}                             & \multicolumn{3}{c}{\cellcolor[HTML]{FFFFFF}\textbf{DS-Java}}     & \multicolumn{3}{c}{\cellcolor[HTML]{FFFFFF}\textbf{DS-Python}}  & \multicolumn{3}{c}{\cellcolor[HTML]{FFFFFF}\textbf{DS-C}}       & \multicolumn{3}{c}{\cellcolor[HTML]{FFFFFF}\textbf{DS-C++}}    \\ \cline{2-13} 
\multicolumn{1}{l}{\multirow{-2}{*}{\cellcolor[HTML]{FFFFFF}Methods}}    & \textbf{AUC}         & \textbf{ACC}        & \textbf{RMSE}       & \textbf{AUC}         & \textbf{ACC}         & \textbf{RMSE}     & \textbf{AUC}         & \textbf{ACC}        & \textbf{RMSE}      & \textbf{AUC}        & \textbf{ACC}        & \textbf{RMSE}      \\ \hline
DisKT                                                                    & 0.5440               & 0.5569              & 0.6191              & 0.5222               & 0.5141               & 0.6436            & 0.5525               & 0.4617              & 0.6987             & 0.5540              & 0.5957              & 0.6079             \\
RouterKT                                                                 & 0.6061               & 0.7098              & 0.4771              & 0.5967               & 0.6989               & 0.4831            & 0.5950               & 0.6913              & 0.4861             & 0.5922              & 0.6839              & 0.4945             \\
csKT                                                                     & 0.6047               & 0.6602              & 0.5057              & 0.5644               & 0.6064               & 0.5383            & 0.5643               & 0.5922              & 0.5526             & 0.5719              & 0.6014              & 0.5422             \\
CL4KT                                                                    & 0.6095               & 0.6723              & 0.5725              &   0.5714               & 0.6211               & 0.6156            &       0.5679               & 0.6044              & 0.6289             &       0.5748              & 0.6330              & 0.6058              \\
sparseKT                                                                 & 0.6095               & 0.6723              & 0.5725              & 0.5714               & 0.6211               & 0.6156            & 0.5679               & 0.6044              & 0.6289             & 0.5748              & 0.6330              & 0.6058             \\
Ours                                                                     & \textbf{0.6882}      & \textbf{0.7682}     & \textbf{0.4073}     & \textbf{0.7012}      & \textbf{0.7418}      & \textbf{0.4172}   & \textbf{0.6852}      & \textbf{0.7195}     & \textbf{0.4334}    & \textbf{0.6855}     & \textbf{0.7387}     & \textbf{0.4315}    \\ \hline
\end{tabular}
\end{table*}

\section{Experiments}
To demonstrate the effectiveness of our proposed model in real-world cross-disciplinary scenarios, we conduct comprehensive experiments in this section. Specifically, we conduct comprehensive experiments on multiple real-world open-source datasets, including cross-disciplinary cold start performance evaluation, ablation experiments, visualization of knowledge state distribution, and parameter analysis.

\begin{table}[t]
\centering
\caption{Summary of Cross-course datasets in PTADisc}
\label{tab:ptadisc_summary}
\begin{tabular}{lcccc}
\toprule
Course & KCs & Exercises & Students & Response Logs \\
\midrule
Python & 685 & 17,787 & 29,454 & 6,454,336 \\
Java & 773 & 16,752 & 29,454 & 4,750,970 \\
C & 847 & 26,056 & 29,454 & 11,378,017 \\
C++ & 547 & 15,172 & 29,454 & 6,587,356 \\
DS & 767 & 21,952 & 29,454 & 7,789,280 \\
\bottomrule
\end{tabular}
\end{table}

\subsection{Experimental setup}
To evaluate the effectiveness of our proposed method, we simulated a variety of cross-disciplinary cold-start scenarios. In this experiment, we employed CL4KT as the pre-trained model and set the contrastive loss to zero. In the pre-training phase, 60\% of the total number of students in the dataset was used as the pre-training set, 20\% as he validation set, and 20\% as the test set. In the training phase, 0.1\% of the total number of students was used as the training set, 20\% as the validation set, and 20\% as the test set. Our method obtains data about overlapping students from the source discipline and the target discipline. In the validation and testing phases, in order to simulate cold-start scenarios, we deliberately hid the interaction records of students in the target discipline, retaining only their scores in the target discipline and relevant data from the source discipline, so as to accurately evaluate the performance of cold-start students in the target discipline.
For computational efficiency, we set the maximum interaction sequence length to 200. We also set the batch size to 64, the learning rate to 0.001, and the maximum number of epochs to 100. To prevent overfitting, we use a dropout layer with a dropout rate of 0.2. We train the model using the Adam optimizer \cite{reyad2023modified}. we adjust the number of expert networks between \{2, 4, 8, 16, 24, 32\}. For the proposed discriminator learning, we adjust the number of types between \{1, 2, 4, 8, 12, 16, 32\}.
To ensure the accuracy of the model evaluation, we keep the parameter settings of the baseline method consistent with those in the original method.

\subsubsection{Dataset}
The cross-course dataset in PTADisc\cite{hu2023ptadisc} contains the learning records of 29,454 students in five different courses, making PTADisc the first dataset that supports cross-disciplinary research and provides new solutions to the cold-start problem in personalized learning. The specific statistical information of the dataset is shown in Table 2.
\subsubsection{Baselines}
For comparison, we use the following baselines.
\begin{itemize}
\item \textbf{CL4KT} \cite{lee2022contrastive}: It proposes a framework for knowledge tracing based on contrastive learning, aiming to learn the relationships between semantically similar or dissimilar samples in learning history.

\item \textbf{sparseKT} \cite{huang2023towards}: It Ranking all attention scores , addressing the common overmatching challenge in attention-based KT models.

\item \textbf{csKT} \cite{bai2025cskt}: It introduces nuclear bias to enhance the model's adaptability to interaction sequences of different lengths. In addition, csKT uses the cone attention mechanism for efficient modeling and performs well even in the case of short interaction sequences.

\item \textbf{RouterKT} \cite{liao2025routerkt}: It proposes a generalized mixture of experts (MOE) approach with multiple concerns as experts and a human-based routing mechanism that significantly improves the performance and reasoning efficiency of existing KT models.

\item \textbf{DisKT} \cite{zhou2025disentangled}: It mitigates cognitive bias by modeling causal effects, differentiating students' abilities to different concepts, and eliminating the confounding effects of students' historical correctness distributions. In addition, an ambivalent attention mechanism is introduced to mask students' ambivalence.
\end{itemize}

\subsection{Performance Comparison}
Table 1 illustrates the overall evaluation results. Our experiment used both AUC \cite{muschelli2020roc}, ACC \cite{gunawardana2009survey} and RMSE \cite{hodson2022root} for overall comparison. From the results, we have the following observations. First, collectively, ACKT achieved the best results in all interdisciplinary scenarios. This is a testament to ACKT's outstanding ability to perform in extreme cold-start environments. Second, in some scenarios, RouterKT, csKT, and CL4KT achieved sub-optimal results in different metrics, respectively. This clearly demonstrates that single-discipline models are ill-suited for cross-disciplinary environments.

\subsection{Ablation Study}
To study the effectiveness of each component of the model, we conducted four variants of ACKT. ACKT-gan refers to removing the preference distribution alignment optimization from ACKT. ACKT-moe refers to using only a single nonlinear mapping function. ACKT-prefer refers to removing the student interaction sequence preferences of the source discipline. ACKT-gate refers to not using a gated network to guide the expert network. The specific results are shown in Table 3. Observations reveal that all variants exhibit varying degrees of performance decline, with ACKT-moe and ACKT-gate showing the most pronounced deterioration. This phenomenon indicates that incorporating appropriately sized cross-disciplinary mediator variables facilitates more accurate characterization of knowledge transfer patterns between disciplines. Concurrently, the gating mechanism plays a pivotal role in assisting expert networks to distinguish students' knowledge state types.

\begin{table}[]

\caption{ACKT ablation experiments in cold-start scenarios from Java→Python}
\label{tab:my-table}
{%
\begin{tabular}{llll}
\hline
Models         & AUC             & ACC             & RMSE            \\ \hline
\textbf{ACKT} & \textbf{0.7065} & \textbf{0.7597} & \textbf{0.4083} \\
ACKT-gan      & 0.6990          & 0.7509          & 0.4151          \\
ACKT-moe      & 0.6937          & 0.7280          & 0.4227          \\
ACKT-prefer   & 0.6996          & 0.7513          & 0.4150          \\
ACKT-gate     & 0.6819          & 0.7278          & 0.4215          \\ \hline
\end{tabular}%
}
\end{table}

\subsection{Impact of different parameters}
In this section, we study the impact of hyperparameters in ACKT. We first analyze the number of experts n, which plays an important role in the CMOE module. Second, we also explore the effect of the number of negative samples NS used for discriminator training and the effect of the loss weight $\lambda$. 

We compared experimental results for different parameters in the java→python transition, as shown in Figure 3(a). We observed that the optimal number of experts is 24. This indicates that a moderate number of experts achieves the best performance, while performance begins to decline when the number of experts is too large. We attribute this to excessive specialization resulting from an excessive number of experts. Figure 3(b) illustrates how model performance varies with different sampling sizes, showing optimal performance when $N_s$ = 16. When the sampling size increases to 32, model performance declines. We attribute this to excessive sampling of non-homogeneous samples, which causes overly biased local features after mapping and consequently leads to suboptimal performance. Figure 3(c) is showing how the model performance changes when $\lambda$ is altered. We observe that when $\lambda\le$ 50, the model performance remains unchanged. This is typically because $\mathcal{L}_{cross}$ dominates absolutely at this point, while the discriminator's capability is too weak to make a difference.

\begin{figure}[h]
	\centering
	\includegraphics[width=\linewidth]{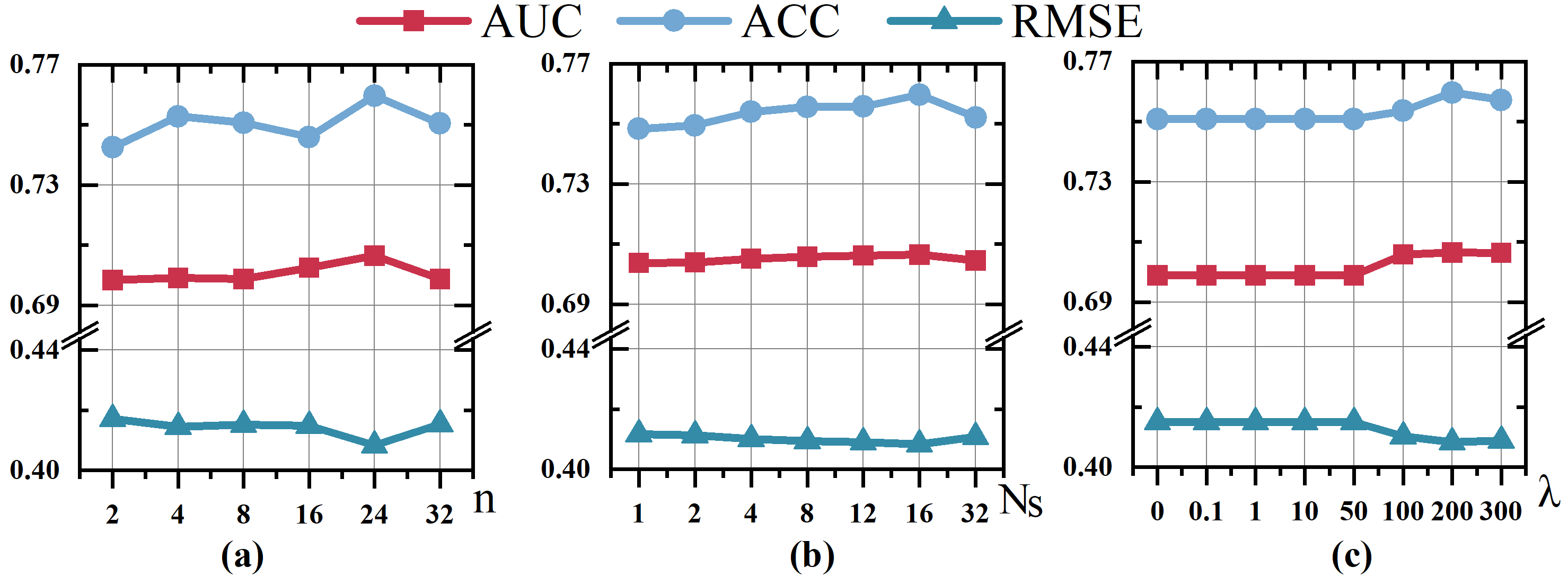}
	\caption{Effect of model parameters on performance}
	\label{FIG:1}
\end{figure}
\begin{figure}[h]
	\centering
	\includegraphics[width=\linewidth]{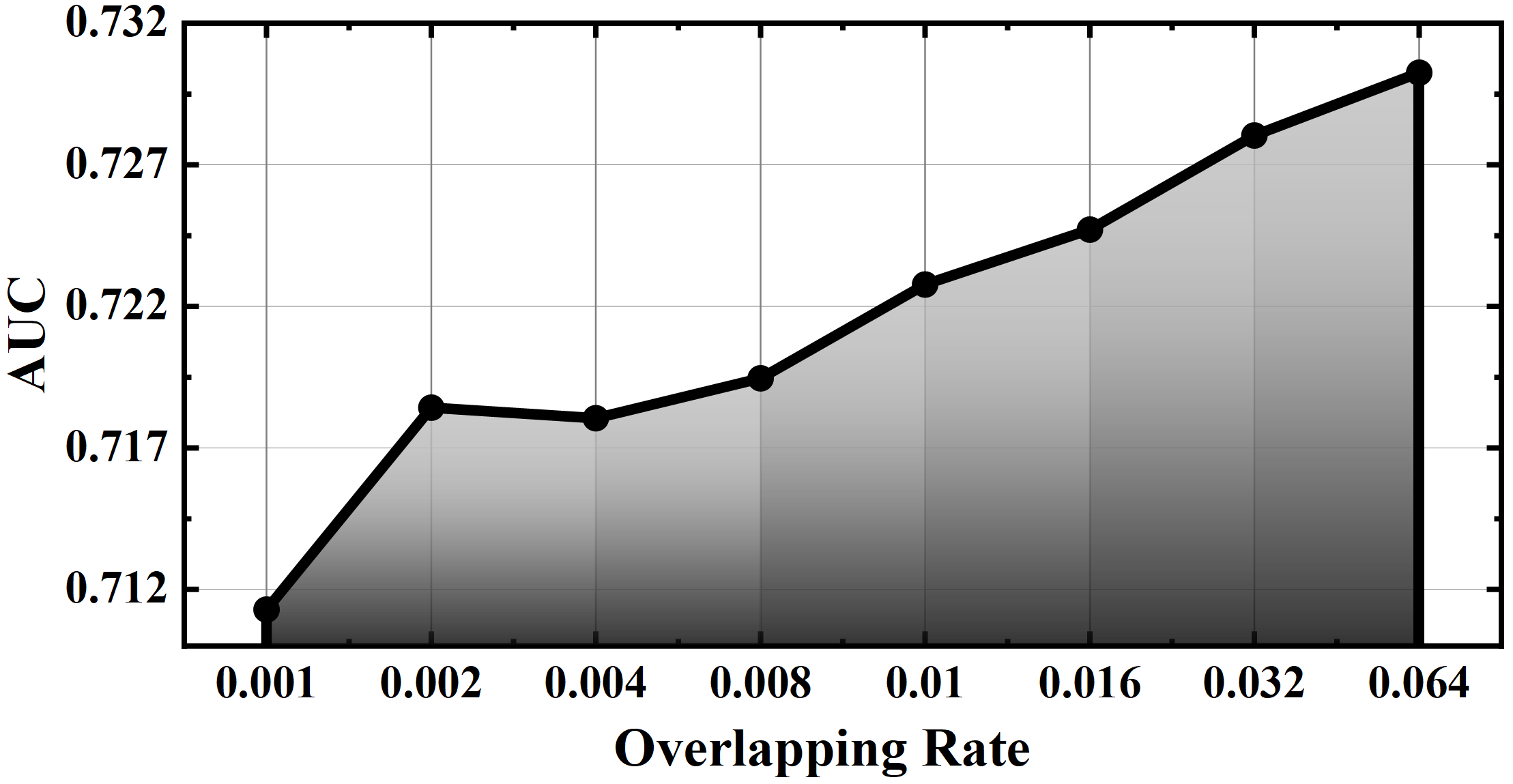}
	\caption{The effect of the number of overlapping students on model performance.}
	\label{FIG:1}
\end{figure}

\begin{table*}[]
\caption{Experimental results on variants}
\label{tab:my-table}
\resizebox{\textwidth}{!}{%
\begin{tabular}{
>{\columncolor[HTML]{FFFFFF}}l 
>{\columncolor[HTML]{FFFFFF}}l 
>{\columncolor[HTML]{FFFFFF}}l 
>{\columncolor[HTML]{FFFFFF}}l 
>{\columncolor[HTML]{FFFFFF}}l 
>{\columncolor[HTML]{FFFFFF}}l 
>{\columncolor[HTML]{FFFFFF}}l 
>{\columncolor[HTML]{FFFFFF}}l 
>{\columncolor[HTML]{FFFFFF}}l 
>{\columncolor[HTML]{FFFFFF}}l 
>{\columncolor[HTML]{FFFFFF}}l 
>{\columncolor[HTML]{FFFFFF}}l 
>{\columncolor[HTML]{FFFFFF}}l }
\hline
\cellcolor[HTML]{FFFFFF}                           & \multicolumn{3}{c}{\cellcolor[HTML]{FFFFFF}\textbf{Java-Python}} & \multicolumn{3}{c}{\cellcolor[HTML]{FFFFFF}\textbf{Java-C}} & \multicolumn{3}{c}{\cellcolor[HTML]{FFFFFF}\textbf{Java-C++}} & \multicolumn{3}{c}{\cellcolor[HTML]{FFFFFF}\textbf{Java-Ds}} \\ \cline{2-13} 
\multirow{-2}{*}{\cellcolor[HTML]{FFFFFF}Variants} & \textbf{AUC}         & \textbf{ACC}         & \textbf{RMSE}      & \textbf{AUC}        & \textbf{ACC}      & \textbf{RMSE}     & \textbf{AUC}        & \textbf{ACC}        & \textbf{RMSE}     & \textbf{AUC}        & \textbf{ACC}      & \textbf{RMSE}      \\ \hline
DisKT*                                             & 0.7018             & \textbf{0.7727}    & \textbf{0.4041}    & 0.6802             & 0.7310           & 0.4564           & 0.6510            & 0.6763              & 0.5277           & 0.6416             & 0.6358          & 0.4615           \\
RouterKT*                                          & \textbf{0.7046}    & 0.7579               & 0.4262            & \textbf{0.6921}    & \textbf{0.7461}   & \textbf{0.4287}   & 0.6640            & \textbf{0.7096}    & \textbf{0.4411}   & \textbf{0.6665}   & \textbf{0.7318}   & 0.4311             \\
csKT*                                              & 0.69812             & 0.7284              & 0.4390              & 0.6839              & 0.6665           & 0.4681             & \textbf{0.6672}    & 0.6726            & 0.4839            & 0.6512             & 0.6762           & \textbf{0.4226}   \\ \hline
\end{tabular}%
}
\end{table*}

\subsection{Performance in non-extreme cold starts}
The motivation for this paper stems from an extreme cross-disciplinary cold-start scenario (i.e., overlap rate = 0.001). To evaluate the model's performance across varying overlap rates, we selected eight distinct overlap rate groups ranging from 0.001 to 0.064 to observe the model's behavior. Figure 4 shows the model performance as varied with the overlap rate. It can be observed that the model's AUC performance generally increases with the rise in overlap rate. This demonstrates the contribution of the proposed method's focus on overlapping regions to cross-disciplinary performance.

\begin{table*}[h]
\centering
\caption{Number of learnable parameters (in megabytes), inference time (in seconds) for different methods in scenario java → python.}
\label{tab:my-table}
\begin{tabular}{cccccccccc}
\hline
Methods         & CL4KT  & SparseKT    & RouterKT & DisKT & csKT & ACKT   \\ \hline
Time(s)  & 8.32  & 13.54       & 0.617    & 3.47 &6.92 
  & 0.03  \\
Parameters(M)  & 1.45  & 16.01   & 1.62   & 3.12 &5.74 
  & 24.39  \\ \hline
\end{tabular}%
\end{table*}

\subsection{Comparison on Variants}
In this section, we replace the KT backbone network to obtain multiple variants of the model , in to verify the generality of the proposed framework. The results are shown in Table 4, where all three variants exhibit a qualitative change in performance during cross-disciplinary cold -start. This demonstrates that our framework not only possesses strong compatibility but also delivers significant performance gains.

\subsection{Visualization}
This section visualizes the effect of the model in cross-disciplinary mapping by visualizing the source discipline, the target discipline, and the knowledge state of the source discipline after mapping through T-SNE \cite{van2008visualizing}. As shown in Figure 5. Here, Cross-moe denotes the approach without the mixed of expert, while Cross-gan indicates the method without considering the generative adversarial loss. Observing the distribution of different colors in the figure reveals that the mapped knowledge state generally aligns more closely with the target discipline's knowledge state. The feature distributions of Cross-moe and Cross-gan are more similar to Source. This fully validates the effectiveness of our proposed interdisciplinary approach and the critical role of the designed components.
\begin{figure}[h]
	\centering
	\includegraphics[width=\linewidth]{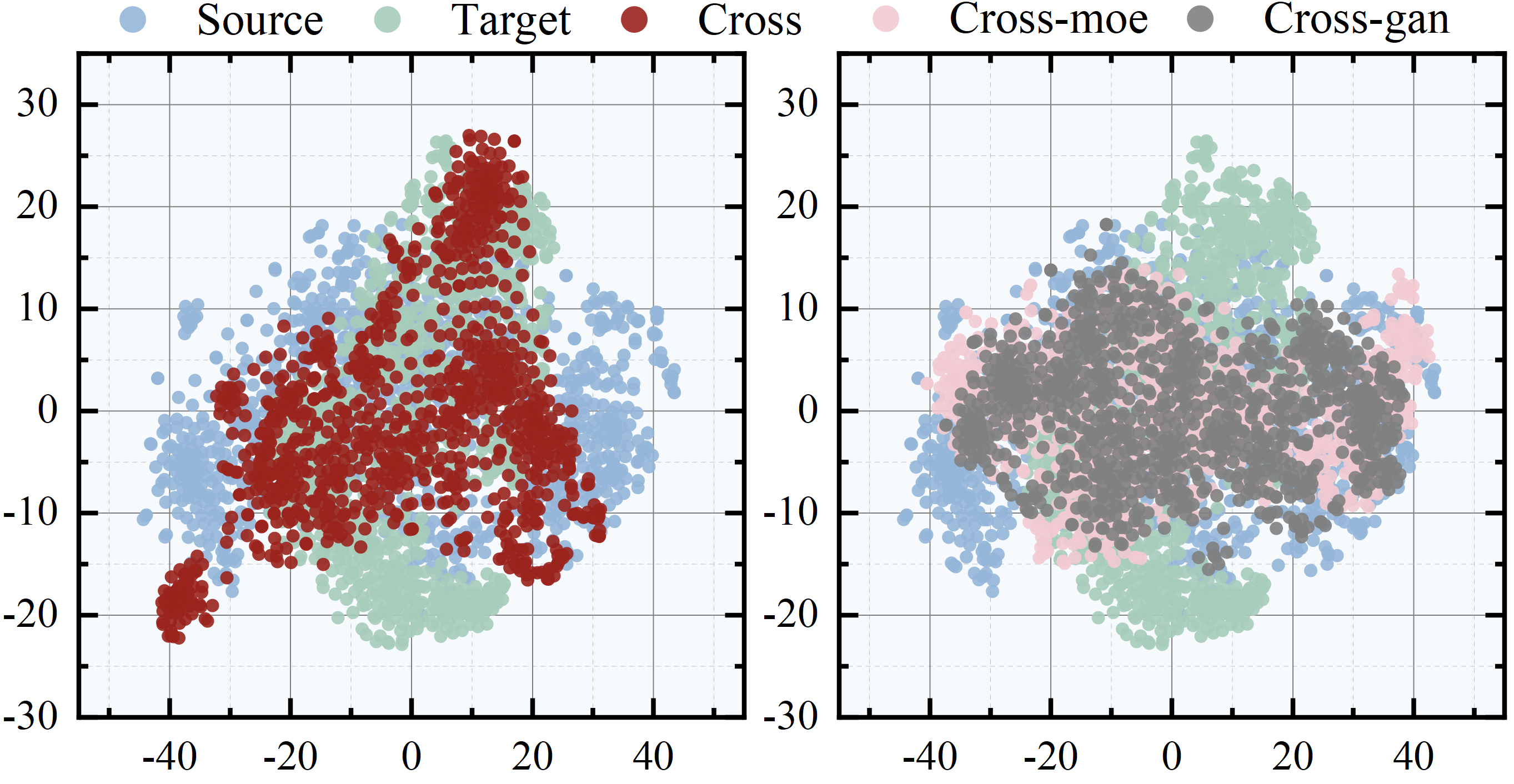}
	\caption{Visualization of students' knowledge states under different circumstances.}
	\label{FIG:1}
\end{figure}

\subsection{Complexity and efficiency analysis}
To systematically evaluate the model complexity and operational efficiency, we accurately count the number of learnable parameters of the proposed method and all baseline models, as well as measure the reasoning elapsed time per round (calendar element) in the Java→Python cold-start scenario with an overlap rate set to 0.1\%.

From the results in Table 5, the learnable parameter count of ACKT reaches 24.39 M, which is about 1.5 folds compared to the next largest baseline SparseKT, while its inference time is only 0.03 s, achieving an order-of-magnitude reduction compared to all the comparison methods. Its ability to maintain low latency despite the significantly larger number of parameters indicates that the framework has high parallel friendliness and hardware utilization, which provides feasibility for subsequent deployment in edge devices or high concurrency scenarios.

\section{Conclusion}
In this paper, we propose a novel ACKT framework focused on cross-disciplinary extreme cold-start knowledge tracing. Specifically, considering that knowledge state transfer relationships across disciplines are often complex , and that the single mapping adopted in current research for cross-disciplinary intermediates is suboptimal, we introduce a novel category-enhanced MOE (CMOE) module to combine common and personalized preference transfer patterns. Furthermore, to mitigate sparse supervision from limited overlapping users, we introduce an unsupervised preference distribution alignment optimization. We propose that preference-similar learners share closer representation distributions after mapping. This approach leverages the core insight that preference-similar learners share closer representation distributions after mapping. The proposed unsupervised optimization strategy enables non-overlapping learners to participate in model training, effectively alleviating performance bottlenecks caused by extreme cold starts. We conduct extensive experiments across 20 cross-disciplinary scenarios spanning five real-world datasets to demonstrate the effectiveness of the proposed ACKT. Although this work reduces the overlap rate constraint to 0.001, it still cannot overcome the limitations imposed by overlapping learners. In future work, we aim to break through the constraints of overlapping learners in cross-disciplinary scenarios and explore cross-disciplinary cold-start knowledge tracing in scenarios without overlapping learners.

\begin{acks}
This work was supported by Yunnan Expert Workstation 202305AF1\-50045;
National Natural Science Foundation of China under Grants 61761045.Yunnan University graduate research innovation fund project No. KC-24249476.
\end{acks}

\bibliographystyle{ACM-Reference-Format}
\bibliography{bibref}










\end{document}